# Non-uniform Fourier Domain Stretching method for ultra-wide-angle wave propagation

TOMASZ KOZACKI,[1] JUAN MARTINEZ-CARRANZA,[1] MAKSYMILIAN CHLIPALA,[1] RAFAL KUKOLOWICZ,[1] AND MONCY S. IDICULA[1]

[1]*Institute of Micromechatronics and Photonics, Warsaw University of Technology, Warsaw, Poland*
*\*tomasz.kozacki@pw.edu.pl*

**Abstract:** Numerical inspection of wide-angle (WA) or ultra-wide-angle (UWA) computer-generated holograms (CGH) is a computationally demanding task. To surpass this limitation, we propose a novel computational approach for fast and accurate WA-CGH reconstruction based on Fast Fourier transform Fresnel diffraction (FrT) and non-uniform frequency hologram magnification. This novel algorithm, referred to as the non-uniform Fourier Domain Stretching (NU-FDS) method, is based on approximating a spherical wave coming from any object point with a parabolic wave, with the points of convergence of the two waves being different. It is supported by a mathematical solution developed using phase-space to determine the frequency distribution needed to find the distribution of non-uniform magnification. It corrects the axial distance of a parabolic wave so the FrT solution can be applied for WA-CGH reconstruction. The NU-FDS algorithm also allows the reconstruction of a partial view with freedom of position and size selection, reducing computation time. In this way, the NU-FDS method enables fast and accurate quantitative assessment of the 3D information coded into the CGH. The presented evidence shows that the NU-FDS algorithm can accurately and efficiently reconstruct large and highly detailed 3D objects from WA and UWA CGH of FoV and resolution up to 120° and 16K, respectively.

## 1. Introduction

Holographic near-eye display (HNED) promises an unparalleled 3D rendering experience for wearable visual systems [1,2]. At the heart of the HNED is digital holography, which allows 3D information to be captured and reproduced using a single hologram. The theoretical capabilities of HNEDs are delivering images with intense brightness [3], rich shape details [4], and, most importantly, reconstruction of 3D scenes that fulfill all the physiological cues of human vision [5–7]. However, current HNED platforms face several restrictions, such as form factor, power efficiency, and slow optoelectronics. Nevertheless, as for any imaging system, one of the most critical problems for HNEDs is generating high-quality content. Such holographic content provides a rich, immersive experience that requires a wide field of view (FoV), covering center (60°) and peripheral (120°) human vision, full parallax, high scene resolution, and even haptic interaction [1]. Therefore, holograms must be created as wide-angle (WA, FoV > 50°) or even ultra-wide-angle (UWA, FoV > 90°). The primary source of 3D content for these devices is computer-generated hologram (CGH) algorithms [8–11], which are calculated for a specific HNED configuration [12,13]. CGH algorithms based on propagation techniques, either one-step [14] or iterative [15,16], cannot create CGH with large FoV. This is because such a CGH requires huge computational resources for wave field propagation, and there is a lack of propagation tools to generate WA-CGH. Nevertheless, several algorithms can calculate WA-CGH or UWA-CGH, which are based on point source approximation [3,10,17] and still need to be researched to deal with features such as shadow and diffusion [6]. Finally, a successful pipeline for generating and displaying images and video in HNEDs requires holographic coding [18–20]. These coding strategies are necessary for optimizing the storage and transmission of information in HNEDs [18]. Hence, the critical element of this holographic pipeline, which includes CGH generation, is the quality control of the reconstructed image [19]. This means it is necessary to develop highly efficient numerical techniques for reconstructing holograms that enable image quality testing and the development of visual quality metrics [18].

To create the required quality control for HNED, there is a strong need to decode wavefront information from WA-CGH or UWA-CGH. For this purpose, it is necessary to develop efficient WA and UWA propagation methods [21]. These propagation algorithms enable quality testing of WA-CGHs [22] or wide-angle holograms of real objects [23]; thus, the quality of WA holograms can be quantitatively addressed. In this way, sources of errors can be systematically corrected. However, propagation techniques become extremely challenging when reconstructing WA-CGHs. This is because the optical field propagates over a large distance, over a large angular span, and for small pixel pitch [24,25]. For this case, basic tools such as angular spectrum (AS) or Rayleigh-Sommerfeld (RS) techniques are inadequate, as they can only reconstruct a very small portion of the scene. Hence, the challenge for WA propagation methods is to generate a full view of the object in a short time, with high accuracy and small computational resources. Consequently, there is a need to develop fundamentally new propagation tools to enable the calculation and reconstruction of WA-CGH and UWA-CGH. With the development of these tools, flexibility in image generation will be available for 3D image relocation and quantitative evaluation of hologram-encoded images, making it possible to test and develop new holographic techniques. For this reason, propagation algorithms that enable the reconstruction of WA-CGHs are an essential research topic. This has been recognized in studies that have proposed several large FoV propagation methods [26–28]. Reference [29] has developed the Multi Fast Fourier Transform angular spectrum method (MFFT-AS) that enhances the process of image reconstruction by using multi-FFT calculations and tile operations in the frequency domain [30]. Reconstruction methods based on the Compact Space Bandwidth Angular Spectrum (CSW-AS), which uses a different output sample size and reduces the space bandwidth product [31], can reconstruct scenes with large FoV [32,33]. The scalable AS method [34] also improves AS performance because, as in CSW-AS, the output pixel pitch is larger than the input pixel pitch. However, due to the discrete nature of the calculation, for which the pixel size is less than 1 µm, the resulting wavefield is subject to sources of error such as aliasing and self-replication [35]. These problems can be solved by applying zero padding schemes in space or frequency domain [36]. Nevertheless, the amount of zero-padding depends on pixel pitch, propagation distance, and target output, among other factors [37–39]. The mentioned factors cause those diffractive calculations to become extremely intensive for pixel sizes below 0.7 µm. Hence, highly efficient propagation algorithms are necessary. Off-axis calculations can provide accurate propagation at large off-axis angles [40], but in the case of WA-CGH, only a very small area of the scene can be reconstructed. Motivated by this need, the Tile-Off-Axis AS (TOA-AS) [28] method for partial image reconstruction was developed to decode large off-axis areas from WA-CGHs. The TOA-AS method combines the tiled, multi-FFT, and CWS-AS approaches but, unfortunately, does not provide a full-sized image. It is worth noting that all mentioned propagation algorithms use uniform sampling in frequency and space. Few research studies have proposed propagation algorithms [41–43] based on non-uniform sampling. Unfortunately, those solutions have not been developed to fulfill the WA-CGHs reconstruction expectations.

In this work, to overcome the current limitations of WA and UWA propagation methods, a new high-performance Non-Uniform Fourier Domain Stretching (NU-FDS) propagation algorithm is proposed. The NU-FDS method is based on the Fast Fourier Transform - Fresnel diffraction (FrT) [36] and non-uniform magnification. The principle of work of this algorithm is based on the idea that the optical field coming from any point of the object covers the entire hologram space but has a small bandwidth. Therefore, such a spherical wave can be approximated by a parabolic wave. However, the convergence points of spherical and parabolic waves are different. The NU-FDS method corrects this discrepancy with a non-uniform magnification applied in the frequency domain. A solution is developed to determine the location of the focus point of the parabolic wave as a result of the applied non-uniform magnification to the hologram. For this purpose, the tools of nonparaxial phase-space (PS) [41], local frequency radius [17], and local frequency position [27] are used. They allow calculating the convergence point of corresponding parabolic waves. On this basis, the paper develops equations to calculate both PS tools and, thus, the convergence point location due to the non-uniform magnification utilized. When using this solution, a distribution of non-uniform magnification is found to correct the reconstruction distance of all paraxial waves, approximating all object waves encoded in the hologram. Thanks to this non-uniform manipulation, the FrT method can be applied to reconstruct WA-CGH. The applied frequency magnification corrects the reconstruction distance. However, to obtain the correct transverse coordinates, non-uniform spatial magnification is also applied at the last step of the algorithm. Hence, the NU-FDS method enables accurate and fast reconstruction of high-resolution WA or UWA CGHs, something that has not been shown until now. The NU-FDS algorithm is flexible when choosing reconstruction areas. Thus, an implementation of the algorithm that allows the reconstruction of a partial view with a selected position and size, further improving the calculation time, is shown. Theoretical analysis, numerical and experimental results are presented to prove the effectiveness and accuracy of the NU-FDS solution, demonstrating its ability to reconstruct such large holograms as 16K and as large FoV as 120°.

## 2. Reconstruction of WA-CGH with Fresnel approximation

The simplified geometry of the HNED is presented in Fig. 1, where hologram and object planes are marked as $(x, y)$ and $(x_o, y_o)$, respectively. For WA-HNED case, the hologram is small, while the reconstructed object is large. The hologram is represented by $N_x \times N_y$ discrete sampling points, which are separated by the pixel pitch $\Delta$. Hologram and object planes are axially separated by distance $z_o$, and thus, image size $B_{xo}$ at $z_o$ is related to pixel pitch as follows

$$B_{xo} = 2z_o \tan\left(\sin^{-1}\left(\frac{\lambda}{2\Delta}\right)\right), \quad (1)$$

where $\lambda$ denotes the wavelength. Equation 1 shows that a small $\Delta$ is needed for a WA hologram. Let's consider the example display parameters $z_o$ = 1 m, $\Delta$ = 0.5 μm, $\lambda$ = 0.5 μm and an 8K (8192 × 4096) hologram. When using these parameters in Eq. 1, it is found that $B_{xo}$ = 1.155 m and FoV = 60°. Thus, for a WA hologram, the ratio between the size of the hologram and the image is a large number, which in the example under consideration for the $x$ direction is $B_{xo}/B_{xh}$ = 398. This large difference between the size of the hologram and the image in practice makes it impossible to use classical methods such as AS or RS, which preserve the sampling rate between the hologram and the object.

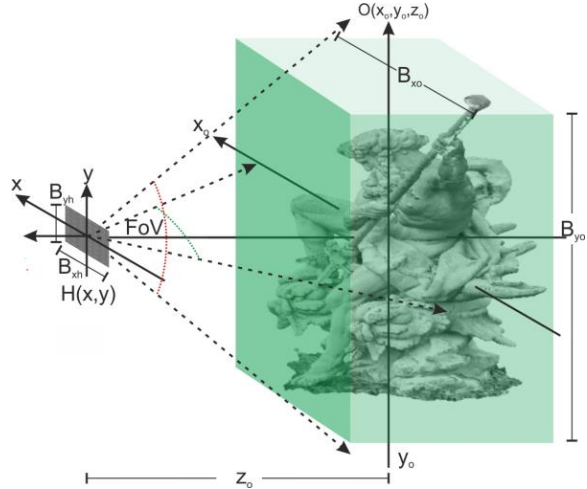

Fig. 1. Geometry for wide-angle hologram reconstruction.

In this work, we reach out to the paraxial FrT method, which has an all-pass transfer function, and the SBP of the signal is maintained during propagation calculation [44]. For the WA hologram, the optical field from any point on the object plane covers an entire hologram space but has a small bandwidth. Later in this Section, we show that a parabolic wavefront can approximate such an object wave in the hologram plane. When considering an object geometry like in Fig. 1, the off-axis points are farther away from the hologram than the on-axis. They are generated by wavefronts of different frequency sizes for both orthogonal directions. Thus, the resolution of the reconstructed off-axis object point is astigmatic and angle-dependent. Accurate calculations require taking such effects into account.

The FrT method is based on the calculation of the integral

$$O(x_o, y_o) = exp\left\{\frac{ik(x_o^2+y_o^2)}{2z_o}\right\} \int H(x,y) \, exp\left\{\frac{ik(x^2+y^2)}{2z_o}\right\} exp\left\{\frac{-ik(xx_o+yy_o)}{z_o}\right\} dxdy, \quad (2)$$

where $k = 2\pi/\lambda$. Using this method, the diffraction field is calculated with a single FFT, and the obtained result is sampled with $\Delta_{ox,y} = \lambda z_o N_{x,y}^{-1} \Delta^{-1}$. As a part of the calculation, the input hologram $H$ is multiplied by a parabolic wave with a focus at distance $z_o$. As a result of this multiplication, all parabolic waves with a focus at $z_o$ are transformed into corresponding plane waves. Then, successively, FFT turns them into the focused points.

The method described in this paper transforms the encoded spherical waves into signals that are an accurate approximation of parabolic waves of the FrT method. This yields a focus point for an off-axis nonparaxial spherical wave with a large angle after processing using Eq. 2. Before developing the WA reconstruction algorithm, let us first analyze the reconstruction errors of the wide-angle hologram using Eq. 2. The evaluation method of the obtained reconstruction errors is based on the PS analysis of the spherical wave

$$H(x,y) = e^{-ikR_o}, \quad (3)$$

where $R_o = \sqrt{(x-x_o)^2+(y-y_o)^2+z_o^2}$ and $(x_o, y_o, z_o)$ is the coordinate of an object point. The propagation method of this work assumes that the hologram given by Eq. 3 can be approximated by a parabolic wave. But this parabolic wave can encode point sources that do not belong to the $z_o$ plane. Thus, the analysis carried out in this Section has two objectives. The first objective, shown in Fig. 2a, is to demonstrate that hologram reconstruction using the FrT algorithm has a large reconstruction error. While the next part, illustrated in Fig. 2d, shows that there exists a point generating parabolic wave that accurately approximates the nonparaxial hologram defined by Eq. 3. Thus, using the appropriate reconstruction distance, the FrT algorithm can give a focused point. Both parts of the error assessment are based on PS analysis of the hologram, where two tools are employed. The first one is a local spatial frequency [33]

$$f_{xl}(x,y) = \frac{1}{2\pi}\frac{\partial ARG\{H\}}{\partial x} = -\frac{x-x_o}{\lambda R_o}, \quad (4)$$

while the second, the local spatial curvature [3]

$$c_{xl}(x,y) = \frac{\lambda}{2\pi}\frac{\partial^2 ARG\{H\}}{\partial x^2} = -\frac{(y-y_o)^2+z_o^2}{R_o^3}. \quad (5)$$

The error of reconstruction of the hologram $H$ with the FrT is shown in Fig. 2b and 2c, where Fig. 2b presents the reconstruction of the hologram of a point object of coordinate [$0.45B_{xo} = 0.52$ m, 0, 1 m]. The result in Fig. 2b illustrates two problems: a large blur and the wrong reconstruction coordinate. This wrong coordinate of the reconstruction can be found from the local spatial frequency evaluated for the center of the hologram, that is $x_{ro} = z_o \lambda f_{xl}(0,0)$. For the example presented in Fig. 2b, $x_{ro} = 0.461$ m. The reconstruction error can be represented as a wavefront error, which is the difference between the exact wavefront and the corresponding paraxial equivalent used in the FrT method. Figure 2c shows the wavefront reconstruction error for the x-direction when reconstructing three points: [0.45$B_{xo}$, 0, 0.5 m], [0.45$B_{xo}$, 0, 1 m], and [0.45$B_{xo}$, 0, 2 m]. The results in Fig. 2c show that the wavefront aberrations, and thus the corresponding reconstruction error, are large. Therefore, FrT cannot be used to reconstruct WA-CGHs. Figure 2c shows that the unacceptable error is also obtained for 4K CGH.

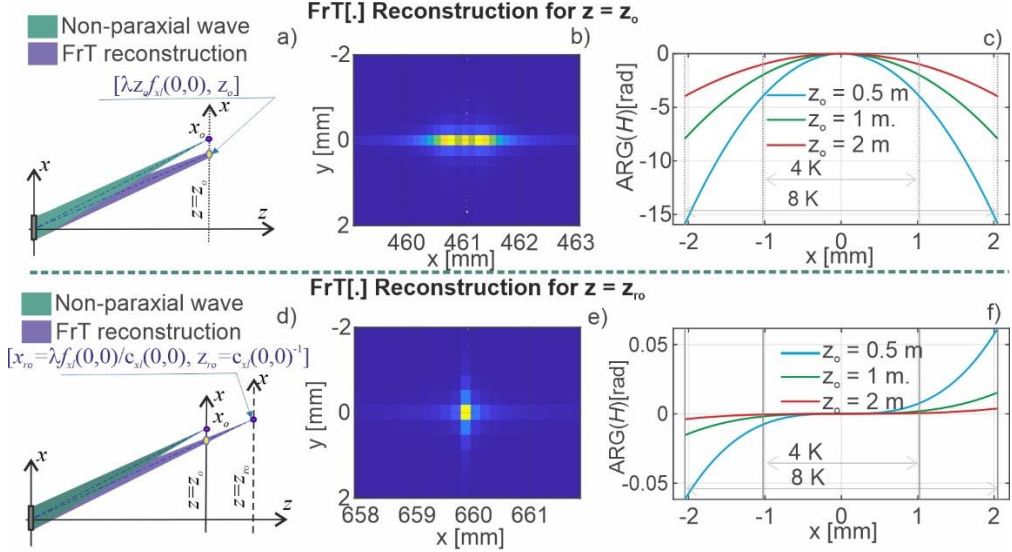

Fig. 2a) Illustration of the errors obtained by FrT reconstructing WA-CGH. b) Reconstruction of 8K CGH of point [0.52m, 0, 1m], c) Wavefront reconstruction errors (cross-section) for three object points [0.45$B_{xo}$, 0, 0.5m], [0.45$B_{xo}$, 0, 1m], and [0.45$B_{xo}$, 0, 2m]. d) Illustration that FrT can give a focused reconstruction of a WA-CGH. e) Reconstruction of 8K CGH of point [0.52m, 0, 1m] for distance $z_{ro}$ = 1.43m. f) Wavefront reconstruction errors for three object points [0.45$B_{xo}$, 0, 0.5 m], [0.45$B_{xo}$, 0, 1 m], and [0.45$B_{xo}$, 0, 2 m] by using $z_{ro}$ = 0.72, 1.43, 2.86 [m].

Above, it was demonstrated that the FrT cannot reconstruct WA holograms. Nevertheless, it is shown here that by choosing a different reconstruction distance $z_{ro}$ an accurate point reconstruction using Eq. 2 can be obtained, as illustrated in Fig. 2d. To determine this distance, we analyze the values of local spatial frequency $f_{xl}(0,0)$ and local spatial curvature $c_{xl}(0,0)$ for the center of the hologram. The direction of reconstruction is determined using the local spatial frequency, while the local spatial position enables the calculation of convergence distance. Thus, using these PS tools, we can find the coordinate [$\lambda f_{xl}(0,0)c_{xl}(0,0)^{-1}$, $c_{xl}(0,0)^{-1}$] of paraxial wave focus, which accurately represents the nonparaxial object wave. Figure 2e presents a hologram reconstruction of object point with [0.45$B_{xo}$ = 0.52 m, 0, 1 m] using Eq. 2 and reconstruction distance $z_{ro}$ = $c_{xl}(0,0)^{-1}$ = 1.43 m. This value is almost 50% larger than $z_o$. It should be noted that the transverse coordinate of the reconstruction is also significantly larger. The point under consideration increases from $x_o$ = 0.52 m to $x_{ro}$ = 0.66 m. This result shows that high-quality reconstruction can be obtained with FrT. Figure 2f illustrates the wavefront reconstruction errors with FrT method when using distance $z_{ro}$ = $c_{xl}(0,0)^{-1}$. A negligible wavefront error is obtained for all considered cases, and the FrT method will be accurate for these distances. Consequently, it can be concluded that by choosing the appropriate reconstruction distance, the FrT method can accurately reconstruct the selected point of the hologram. However, reconstructing the entire hologram is impossible because the reconstruction distance for each point is different.

## 3. Non-uniform hologram magnification in Fourier space

The propagation method proposed in this work is based on the manipulation of spherical wave parameters, such as the 3D position of the reconstruction, so the reconstruction point using FrT is shifted to the location of the object point. The manipulation by which the 3D position of the holographic image is corrected is done by magnifying the hologram. The magnification is realized by stretching the hologram. However, the well-known hologram stretching technique cannot be used for two reasons: firstly, the stretching method is limited to paraxial holograms, and secondly, it is a global operation, i.e., applied to the entire hologram. Therefore, when using global stretching, only the reconstruction of a single point will be corrected.

In contrast, in this work, the hologram stretching technique must be nonparaxial, applied in the frequency domain, and, most importantly, with a non-uniform distribution. The Fourier spectrum of the hologram given by Eq. 3 is

$$\tilde{H}(f_x, f_y) = e^{i\phi_o} = e^{2\pi i f_z z_o + 2\pi i f_x x_o + 2\pi i f_y y_o}, \quad (6)$$

where $f_z = \sqrt{\lambda^{-2} - f_x^2 - f_y^2}$. This paper analyzes the frequency representation of a hologram i.e., the local frequencies. Thus, the local frequency parameters of the hologram can be related to the local spatial information. For this purpose, PS analysis is used, linking the spatial parameters of the image to its spectrum. The most common application is PS in space, using Eqs. 4 and 5. In this work the PS analysis is performed based on the spectrum of the hologram, and thus, the appropriate tools must be introduced. The development is presented for frequency $f_x$. For the frequency component $f_y$ the analysis is the same, the variables are exchanged only.

The first tool is known and is referred to as local frequency position [30]

$$x_l(f_x, f_y) = \frac{1}{2\pi} \frac{\partial \phi_o}{\partial f_x} = -\frac{z_o \lambda f_x}{f_z} + x_o, \quad (7)$$

where $\phi = ARG\{h\}$ given by Eq. 6. The second tool, local frequency radius, is introduced in [17] and it is defined as

$$r_{lx}(f_x, f_y) = \frac{1}{2\pi} \frac{1}{\lambda} \frac{\partial^2 \phi_o}{\partial f_x^2} = \frac{(1-\lambda^2 f_y^2) z_o}{f_z^3}. \quad (8)$$

The local frequency radius is equivalent to the local spatial curvature described in Eq. (5) [3] but enables calculations in the frequency domain. Using these tools, it is possible to find the parameters of the spherical wave encoded in the hologram by analyzing its spectrum. Here, the non-uniform hologram magnification is based on the analysis of hologram frequency components given by Eq. 6 using

$$f_x' = f_x m_x(f_x, f_y), \quad (9)$$

$$f_y' = f_y m_y(f_x, f_y), \quad (10)$$

where $m_x$ and $m_y$ define applied non-uniform frequency hologram stretching for frequencies $f_x$ and $f_y$, respectively. When applying this non-uniform stretching to the hologram, the local frequency position and local frequency radius become

$$x_l'(f_x, f_y) = x_l(f_x, f_y) \left[\frac{\partial f_x'}{\partial f_x}\right]^{-1}, \quad (11)$$

$$r_{lx}'(f_x, f_y) = r_{lx}(f_x, f_y) \left[\frac{\partial f_x'}{\partial f_x}\right]^{-2} + \frac{x_l(f_x, f_y)}{\lambda} \left[\frac{\partial^2 f_x'}{\partial f_x^2}\right]. \quad (12)$$

These equations show that when non-uniform stretching is used, the values of the local frequency position and local frequency radius are modified by the derivatives of applied magnification. Therefore, Eqs. 11 and 12 enable relating the 3D positions of the point sources, encoded in the hologram, before and after non-uniform magnification. To evaluate these coordinates, these equations are analyzed for $f_{x0} = x_o/\lambda/(x_o^2 + y_o^2 + z_o^2)^{\frac{1}{2}}$, and $f_{y0} = y_o/\lambda/(x_o^2 + y_o^2 + z_o^2)^{\frac{1}{2}}$, which is the frequency generated by the point source at the center of the hologram. The second term in Eq. 12 vanishes for this frequency since the local frequency position equals zero. Consequently, Eq. 12 is simplified to

$$r'_{lx}{}^{(f_{x0},f_{y0})} = \frac{(1-\lambda^2 f_{y0}^2)z_o}{(1-\lambda^2 f_{x0}^2-\lambda^2 f_{y0}^2)^{\frac{3}{2}}}\left[\frac{\partial f_{x'}}{\partial f_x}\right]^{-2}. \quad (13)$$

## 4. Non-uniform Fourier Domains Stretching method for wide-angle hologram reconstruction

The developed NU-FDS method is based on Eqs. 2 and 13. Consider again reconstruction of nonparaxial hologram using Eq. 2. As shown in Fig. 2b, the FrT method produces large errors. This is because, for the frequency $f_{x0}$ associated with the object point $x_o$, the local frequency radius differs significantly from $z_o$. The proposed method corrects this error by applying a non-uniform magnification for each object point and, thus, for each frequency of the hologram. Hence, the condition for finding magnification is calculated as

$$r'_{lx}(f_x, f_y) = z_o, \quad (14)$$

which yields

$$\frac{\partial f_{x'}(f_x,f_y)}{\partial f_x} = \sqrt{1-\lambda^2 f_y^2}\left(1-\lambda^2 f_y^2 - \lambda^2 f_x^2\right)^{-3/4}. \quad (15)$$

Integrating this equation for $f_x$ and inserting the result into Eq. 11, the formula describing the desired non-uniform magnification is obtained and is given by

$$m_x(f_x, f_y) = \left(1-\lambda^2 f_y^2\right)^{-\frac{1}{4}} {}_2F_1\left(\frac{1}{2},\frac{3}{4};\frac{3}{2};\frac{\lambda^2 f_x^2}{1-\lambda^2 f_y^2}\right), \quad (16)$$

where $_2F_1$ is the hypergeometrical function [45]; the same procedure is carried out for $m_y$. The frequency-dependent magnification defines the required non-uniform stretching of the FT of a hologram, which is introduced according to Eqs. 9 and 10. The result is a sharp point of a reconstruction. However, when this stretching is applied, the reconstruction of the hologram using Eq. 3 is obtained at the point ($x_o'$, $y_o'$), which can be estimated using the following expression

$$x'_o = z_o \lambda f_x m_x(f_x, f_y), \quad (17)$$

$$y'_o = z_o \lambda f_y m_y(f_x, f_y). \quad (18)$$

The NU-FDS algorithm consists of five steps: two of those initialize the non-uniform magnification, and three for processing the hologram, which are illustrated in the algorithm block diagram shown in Fig. 3. The algorithm begins with initialization step 1 in which non-uniform magnifications $m_x$ and $m_y$ are found according to Eq. 16. The calculation time associated with the non-uniform magnification, according to Eq. 16, is too large. For example, calculating $m_x$ for hologram 128×128 requires 16.5 s. Therefore, a direct method of determining magnification cannot be used. To increase the computational speed of this element of the algorithm, a one-dimensional polynomial is fitted to a small number of samples, which are determined by Eq. 16. In the implementation of the algorithm, the 8-degree polynomial is fitted to 51 frequencies, which are uniformly distributed across the bandwidth of the hologram. The resulting polynomial coefficients are used as a substitute for the hypergeometrical function and are applied according to Eq. 16. By comparison, the calculation for the approximate method for a 4K hologram takes 0.89 s. An example of non-uniform magnification for mapping the frequency of an 8K hologram is shown in Figure 4a. Presented non-uniform magnification is greater than one. Thus, the spectral components of the hologram are shifted towards higher frequencies. For example, component at $\mathbf{f_x} = [0.75\ \mu m^{-1}, 0]$ is casted to frequency location $\mathbf{f_x'} = [0.8\ \mu m^{-1}, 0]$.

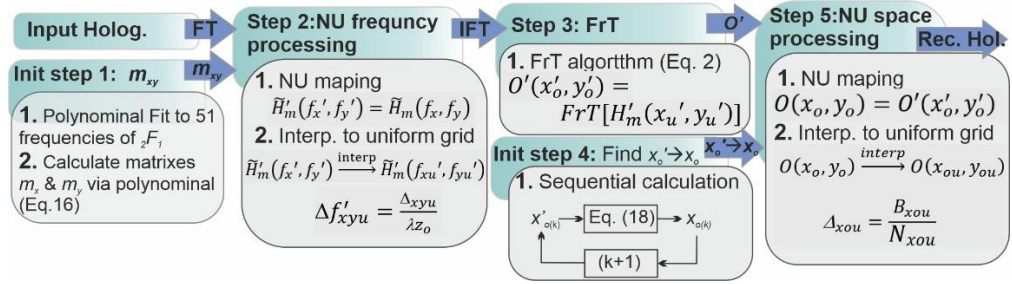

Fig.3. Algorithm flowchart for the NU-FDS method.

The computational step two of NU-FDS algorithm performs a frequency non-uniform mapping that is applied directly to the hologram spectrum as $\widetilde{H}_m'(f_x', f_y') = \widetilde{H}(f_x, f_y)$, and later to the result, interpolation to the uniform frequency grid $(f_{xu}', f_{yu}')$ is performed, where Eqs. 9 and 10 are applied. This step gives uniformly distributed spectrum $\widetilde{H}_m'(f_{xu}', f_{yu}')$. The result is inverse Fourier transformed. The step three is FrT given by Eq. 2, which gives $O'(x_o', y_o')$. The result $O'$, is processed in the step five, which performs distortion correction of the wavefield, similarly to step two, there is a non-uniform mapping according to $O(x_o, y_o) = O'(x_o', y_o')$, where $(x_o, y_o)$ is the correct spatial position of the object point, and successively the result is interpolated from the non-uniform $(x_o, y_o)$ to uniform grid $(x_{ou}, y_{ou})$. The parameters needed to correct the distortion in step five are found in initialization step four. The parameters of non-uniform output spatial mapping are also calculated using one-dimensional polynomial fitting coefficients. Unfortunately, Eq. 17 and 18 cannot be applied directly because they refer to $x_o \rightarrow x_o'$, and not $x_o' \rightarrow x_o$, as needed. Therefore, three interpolations are required: two to find the spatial coordinates and one for the output result. For a 4K hologram, for example, these calculations take 57.3 seconds. To reduce the computational time of this part of the algorithm, we use sequential calculations with four steps, which find the mapping $x_o' \rightarrow x_o$ using

$$x_{o(k)} = \frac{x_o' \sqrt{x_{o(k-1)}^2 + y_{o(k-1)}^2 + z^2}}{z m_x(f_{x0(k-1)}, f_{y0(k-1)})}, (18)$$

where in the first step $x_{o(1)} = x_o'$. In comparison, in the case of a 4K hologram, the calculation time is reduced to 30.6 seconds. The convergence error for step 4 of the algorithm for an 8K hologram and $z_o = 1$ m is illustrated in Fig. 4c for the full matrix $\mathbf{X_o}$. The error shown is normalized to the sample size of the output result as $(x_{o(4)} - x_{o(5)})/\Delta_{xo}$. Figure 4c shows that for four steps, the maximum error of the coordinate calculation is below half a pixel. Figure 4b presents an example of the non-uniform output spatial magnification $x_o'/x_o$. It can be seen that $x_o'$ is always less than $x_o$. Thus, the output spatial non-uniform magnification is smaller than one. For example, for object coordinate $\mathbf{x_o} = [0.75\ m, 0]$ $\mathbf{x_o'} = [0.67\ m, 0]$. NU-FDS is based on data interpolation, so the output sample size and the number of output samples can be freely chosen. Therefore, we start by selecting the parameters of the output result, as $[B_{xo}, B_{yo}]$ and $[N_{xo}, N_{yo}]$. Most often, it is desirable to reconstruct the entire holographic image. Thus $B_{xyou} = B_{xyo}$, which is defined using Eq. 1. The output selection of sampling means that $\Delta_{xyou} = B_{xyo}/N_{xyo}$. Within the algorithm frequency grid $(f_{xu}', f_{yu}')$ is related by the FrT algorithm with the $(x_o, y_o)$, therefore, the frequency sample is given by $\Delta f_{xyu}' = \Delta_{xyu}/\lambda z_o$.

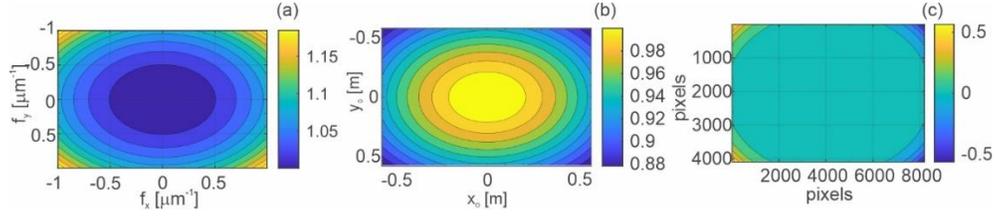

Fig. 4. Example of non-uniform magnifications for 8K hologram for $z_o = 1$ m: a) for stretching FT of hologram ($m_x(f_x, f_y)$) and b) for stretching the output result $x'_o/x_o$. c) Convergence error of step nr 4 of mapping $x_o'$ → $x_o$ normalized to the pixel size ($x_{o(4)} - x_{o(5)})/\Delta_{xo}$.

The NU-FDS algorithm was employed to reconstruct the hologram of three point-source objects with coordinates [0.45$B_{xo}$, 0, 0.5 m], [0.45$B_{xo}$, 0, 1 m], and [0.45$B_{xo}$, 0, 2 m], respectively. The corresponding wavefronts are encoded in an 8K hologram for three different pixel pitches, $\Delta = 0.75, 0.5, 0.3$ [µm] and $\lambda = 0.5$ µm, which corresponds to FoV = 39°, 60°, 113°, respectively. Notably, the examined FoV is very large. Fig. 5a-5c shows the wavefront reconstruction errors for the x-direction for the three-point hologram. The plots show that the largest wavefront reconstruction error is $max|W(x, y)| = 0.11\lambda$. The Rayleigh wavefront criterion [46] states that for $max|W(x, y)| < \lambda/4$, the maximum intensity of the reconstruction point should be at least 80% of the ideal case. Thus, the aberration obtained for the worst case, shown in Fig. 5, is 2.3 times smaller than $\lambda/4$. This result confirms that the wavefront aberration is small. Therefore, it can be considered that the algorithm gives a negligibly small wavefront error, and thus, the result of the algorithm is accurate. Figure 5 shows an interesting feature of the algorithm: the dependence of the wavefront error on the pixel pitch. The accuracy of the NU-FDS method is high for wide angles (small pixels), but for lower angles (large pixels), it starts to fail. The method should be used for holograms with pixels than 1 µm. For the object point [0.45$B_{xo}$, 0, 0.5 m] and $\Delta = 1$ µm NU-FDS still satisfies Rayleigh criterion $max|W(x, y)| < \lambda/4$. Moreover, Fig. 5 shows that the data has reduced spatial support of the related wavefront after spatial interpolation. This is an expected result since the resolution of the reconstruction decreases as a cosine of the angular coordinate of a given point.

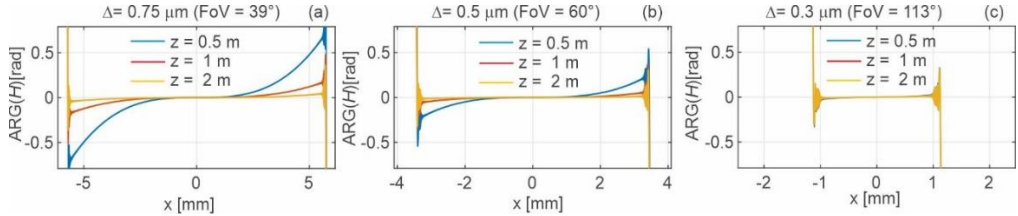

Fig. 5. Wavefront reconstruction errors for 8K hologram reconstruction using NU-FDS method for three object points [0.45$B_{xo}$, 0, 0.5 m], [0.45$B_{xo}$, 0, 1 m] and [0.45$B_{xo}$, 0, 2 m] for three different pixel pitches: a) $\Delta = 0.75$ µm, b) $\Delta = 0.5$ µm, and c) $\Delta = 0.3$ µm.

The developed method is flexible in terms of choosing the reconstruction area. Moreover, this work proposes implementing the algorithm that allows reconstruction of a partial view with a selected position and size. This allows more efficient hologram testing. The method is based on a simple modification of the NU-FDS algorithm, which involves selecting a portion of the hologram spectrum processed in the algorithm. For simplicity of implementation, we consider limiting the bandwidth to rectangular frequency regions. For example, to reconstruct the right top corner of the hologram, we process frequency $f_x > 0$ and $f_y > 0$. However, the experimental Section shows a more interesting implementation of the partial-view reconstruction method. The proposed algorithm processes hologram frequencies: $-¼\Delta^{-1} > f_x > ¼\Delta^{-1}$ and $f_y > 0$. We reconstruct the TOP-CENTER view from the hologram by selecting such a frequency range.

This implementation is shown in the experimental Section. Hence, the presented algorithm is flexible regarding the choice of position and size of the reconstruction area. The computation time of the developed algorithm for full image and partial reconstruction is shown in Table 1, where, as expected, the computation time for partial reconstruction is about four times faster. The calculations were performed on a PC with CPU i7-11850H @ 2.50GHz equipped with 64 GB RAM.

**Table 1: Computational speed of the NU-FDS method**

| Reconstruction method | Time (2K) | Time (4K) | Time (8K) |
|---|---|---|---|
| Full FoV | 23.2 s | 105.3 s | 492.8 s |
| Partial (½FoV$_x$×½FoV$_y$) | 5.7 s | 28.9 s | 150.2 s |

To demonstrate the limitations of current algorithms, we tested the two best approaches designed to improve WA computing, namely MFFT-AS [26] and CSW-AS [27] methods. The resulting hologram reconstruction times for three values of pixel pitches and three resolutions are shown in Fig. 6. For the smallest pixel size, which corresponds to FoV 38°, MFFT-AS required 410 min. to compute the reconstruction of 4K CGH while CSW-AS needed over 1 min. We also tried reconstructing the hologram for $\Delta = 0.5$ μm (FoV 60°) and 4K. In this case, both algorithms failed. MFFT-AS did not complete the calculation in three days, while CSW-AS required matrixes of sizes 110000 x 110000, which is impossible for most computers.

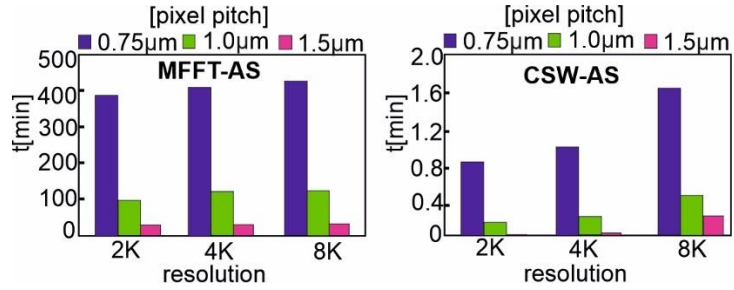

Fig. 6. Calculation times of reconstruction of CGH for different pixel pitches and resolutions: a) MFFT-AS, b) CSW-AS.

## 5. Simulations and Experimental Results

In this Section, the numerical and experimental verification of the NU-FDS method was carried out, which can be divided into four parts. The first part illustrates the WA-CGH reconstruction of a simple test pattern object; the second part shows the CGH reconstruction that encodes the 3D object of complex shape, the Neptune statue, and the third part presents the experimental data. The fourth part is a reconstruction of ultra-wide-angle CGH, showing that the method can reconstruct very large objects; also, for this CGH, for comparison, FrT reconstruction is presented.

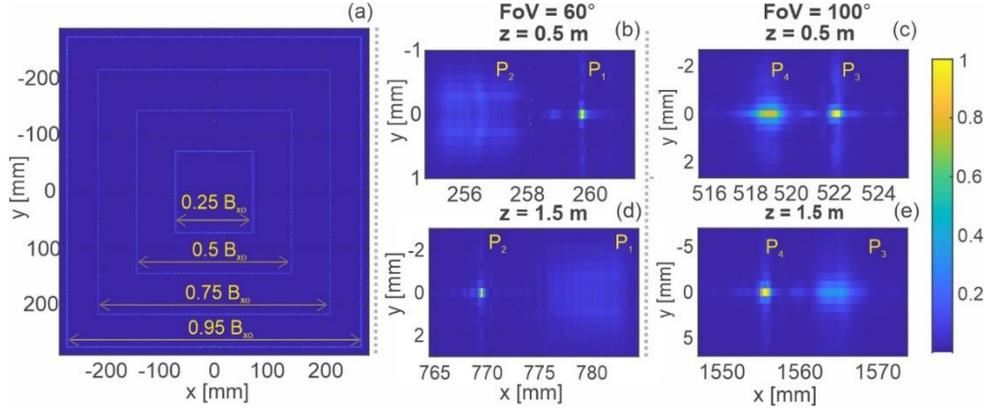

Fig. 7. Reconstruction of point object holograms: (a) CGH1 (FoV = 60°): the object of set squares at 0.5 m. (b), (d) CGH2 (FoV = 60°): object of two points $P_1(0.45B_{xo}, 0, 0.5)$, $P_2(0.45B_{xo}-0.01m, 0, 1.5m)$, focus at $P_1$ and at $P_2$; (c), (e) CGH3 (FoV = 100°): object of two points $P_3(0.45B_{xo}, 0, 0.5)$, $P_4(0.45B_{xo}-0.01m, 0, 1.5m)$, focus at $P_3$ and at $P_4$.

Figure 7 illustrates the results of the first numerical experiment, where the reconstructions of test holograms are presented for CGH of 60° and 100°. The first object, which has a complex shape, proves the correct reconstruction of the entire FoV, while object 2 validates the detailed reconstruction. Figure 7a presents the reconstruction of a flat object, a set of squares of different sizes. The object is located at 0.5 m, and the dimensions of the squares are adjusted to the size of $B_{xo}$ = 577.4 mm. The result shows that the shape is well reconstructed. However, the small features are not well seen due to the object's large size. Thus, in this experiment, the holograms CGH2 and CGH3 of 60° and 100° FoV are generated for single isolated points located at the edge of hologram FoV, for which reconstruction is shown in Fig. 7b-e. CGH2 is generated for two points, one at $z$ = 0.5 m $P_1(0.45B_{xo}, 0, 0.5$ m$)$ and one at $z$ = 1.5 m, $P_2(0.45B_{xo}-0.01$ m, 0, 1.5 m$)$. The point $P_2$ is shifted in $x$ by 10 mm to prevent overlapping. Figures 7b and 7d show the sharp reconstructions of the points at $z$ = 0.5 m and 1.5 m, respectively. This test also shows algorithm capabilities to reconstruct larger FoV = 100°. CGH3, with Δ=0.33 μm, thus FoV = 100°, was designed for two points: $P_3(0.45B_{xo}, 0, 0.5$ m$)$, $P_4(0.45B_{xo}-0.01$ m, 0, 1.5 m$)$. Figures 7c and 7e show sharp reconstructions. It is worth noting that the resolution for the x-direction is reduced, which is the expected result since it decreases according to the cosine of the angular coordinate. See also Fig. 5c.

The second numerical experiment is connected with the optical reconstruction of a CGH in an HNED, for which details are presented below. The numerical and optical tests are performed for the same 3D point set; however, in each case, a different hologram is calculated because the optics of the experimental display cuts the FoV, which is available according to the pixel pitch of the computed hologram. The size of the 3D object is chosen to fit the parameters of the HNED, where the object's height is selected to occupy the entire FoV of the display. In this experiment, the CGH hologram was calculated using the Fourier Domain Method (FDM) [17] for a 3D point cloud with 4.2 million points representing a statue of Neptune. The dimensions of the object are 1.19 m (width) × 1.82 m (height) × 0.98 m (depth).

The parameters of the CGH computed for the numerical test shown in Fig. 8 are 8K, λ = 0.64 μm, Δ = 0.5 μm. Thus, the corresponding FoV = 79.6°. The left image in Fig. 8 presents the full reconstruction of the object at $z_{REC}$ = 0.85 m. At this depth, the trident blades are in focus. The image was also reconstructed for a distance of 1.52 m, where the right hand of the Neptune is in focus. The corresponding focus images are shown on the insets at the right part of Fig. 8. To improve the quality of the image, speckle averaging was applied. Ten CGHs were calculated using a random phase and then averaged in the display. The figure on the right shows the result of the CGH reconstruction using a partial reconstruction scheme. In this case, the top–center view was reconstructed.

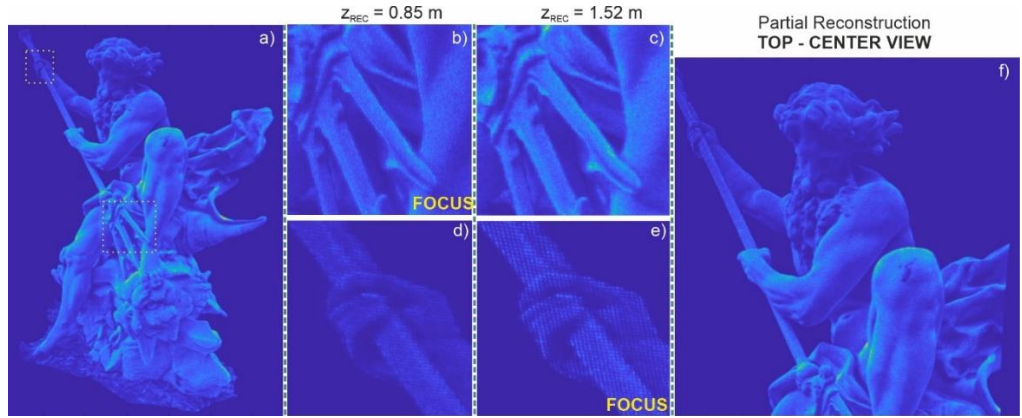

Fig. 8. Numerical reconstruction of 8K CGH of 3D points: a) full reconstruction of the hologram, b) - e) zooms from different focus reconstructions, f) reconstruction computed by using a partial reconstruction scheme.

In the third experiment, the CGH of the statue of Neptune is optically reconstructed in the HNED schematically illustrated in Fig. 9a. The display is built in a pupil-forming architecture [47]. It consists of an illumination module, 4K SLM (HoloEye GAEA 2.0, pixel pitch 3.74 μm, pixel count 4160×2464), and a 4F imaging setup. The laser beam ($\lambda = 640$ nm) focuses on the illumination module's pinhole plate. Behind the filter, a quasi-point source is formed in the focal plane of the collimating lens $L_c$. Using a beam-splitter, the SLM is illuminated normally by the plane wave. Next, light diffracted by the SLM propagates through a 4F imaging setup composed of lenses $L_1$ with $F_1 = 200$ mm and $L_2$ with $F_2 = 21$ mm. Lens $L_2$ is Meade Series 5000 Mega Wide-Angle Eyepiece with 100° FoV. In the back focal plane of the lens $L_2$ eyebox is formed, which is optically conjugated with the SLM plane. The magnification of the setup equals 0.105, and thus, the pixel pitch at the eyebox plane is 0.39 μm. This gives a maximum FoV equal to 110°. However, such a large angle is not supported by the optics employed in the display. Smartphone's large FoV camera cuts part of the object used to capture the reconstructed hologram. The result of the optical reconstruction of CGH calculated for display parameters is shown in Fig. 9b. For better visualization of the details, two regions of the reconstructed image are enlarged, which are shown in Fig. 9c and Fig. 9d. Chosen areas correspond to these presented for numerical reconstructions in Fig. 8. It can be seen that all of the object details are visible. However, in the employed display, we are unable to illustrate the difference between defocusing at different depth planes.

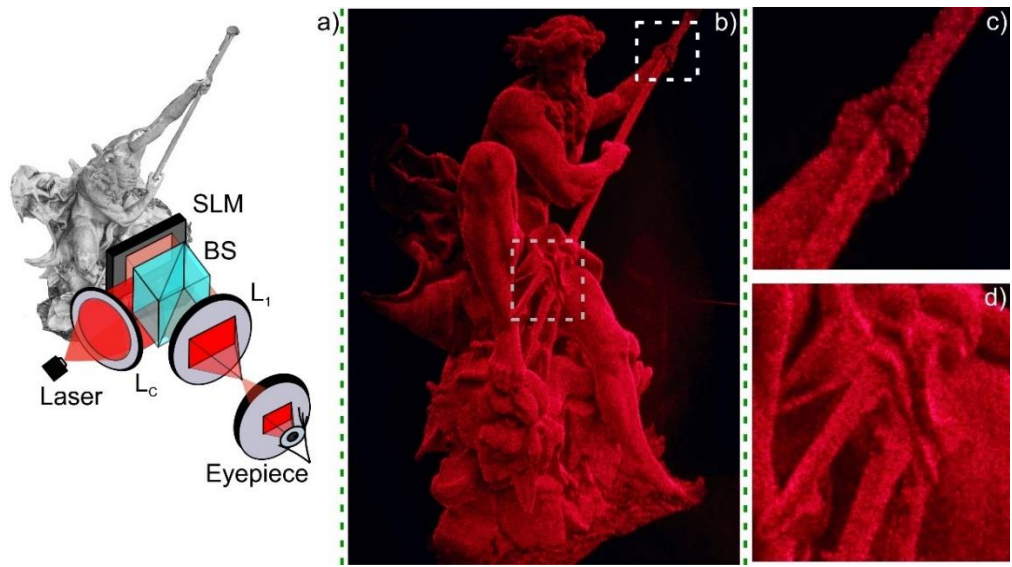

Fig. 9. a) Scheme of near-eye holographic display. b) Optical reconstruction of Neptune statue, c)-d) zooms for different object's parts.

The final experiment presents the numerical reconstruction of ultra-wide-angle CGH. The hologram has a pixel pitch of 0.29 μm and 16K resolution. This pixel pitch corresponds to FoV = 120°. The CGH was again generated using FDM for the 3D cloud of the Neptune statue; this time, it consisted of 2.6 million points after occlusion. For this visualization, the dimensions of the object are 3.76 m (width), 5.37 m (height), and 2.56 m (depth), where the front of the object is at a distance of 0.6 m. Figure 10 presents the obtained results. Figure 10a shows the full reconstruction obtained with NU-FDS for a depth of 0.8 m. This time, no speckle averaging was applied. Figure 10c shows a zoom of the full reconstruction. To present the effect of focusing in Figure 10c, additional zoom is included to show the fringes resulting from the non-uniform fringe-like distribution of points in the cloud. Figure 10c shows the focus of the front part of the object; to also show the focus of the part of the object located at a further distance, the hologram was reconstructed via NU-FDS for a depth of 2.4 m, and the zoom in Fig. 10d is a small part of this reconstruction. This figure shows fringes similar to those in Fig. 10c. Figure 10b shows a reconstruction of this demanding hologram using FrT. Strong 3D deformations are visible; an unnatural view was obtained. One of the visible deformations is the apparent bending of the trident handle. Zooms in Fig. 10e and 10f of the reconstruction obtained using FrT show deformations and blurring, demonstrating an incorrect depth reconstruction using FrT. The zooms in Fig. 10e and 10f contain areas of the reconstructed object similar to those shown in Fig. 10c and 10d and are calculated for the same reconstruction distances.

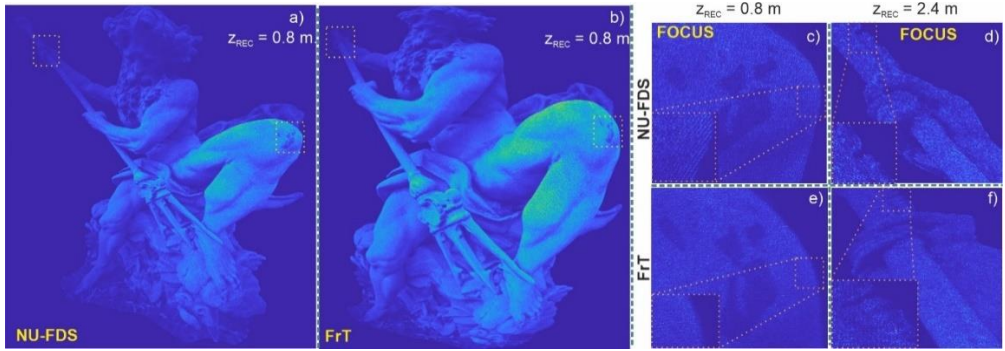

Fig. 10. Numerical reconstruction of 16K CGH of FoV = 120° of 3D points: a) full reconstruction using NU-FDS for depth 0.8 m, b) full reconstruction using FrT for depth 0.8 m, c) - d) zooms at focus via NU-FDS, c) - d) zooms at focus via FrT.

## 6. Conclusions

This paper presents the NU-FDS propagation method based on FrT solution and non-uniform frequency hologram magnification, which is capable of fast and accurate reconstruction of high-resolution WA-CGH. Theoretical analysis, numerical and experimental results prove the effectiveness and accuracy of our NU-FDS, demonstrating its ability to reconstruct holograms of FoV and resolution up to 120° and 16K, respectively. The developed algorithm makes it possible to reconstruct a partial view of the object with freedom of position and size selection, reducing computation time. For example, it takes 105 seconds to reconstruct a 4K hologram, compared to 29 seconds for a partial view (½FoV$_x$ × ½FoV$_y$).

It is important to note that NU-FDS approximates object spherical waves by parabolic waves. However, the convergence points of spherical and parabolic waves are different. Therefore, FrT cannot be directly applied to the reconstruction of WA-CGH. Thus, non-uniform magnification used in the frequency domain corrects the axial distance of a parabolic wave so the FrT can be applied for the reconstruction of WA-CGH. A theoretical solution is developed to determine the location of the focus point of the parabolic wave as a result of the applied non-uniform magnification to the hologram. To this end, the local frequency radius and local frequency position are applied to calculate the focus point of a parabolic wave due to applied non-uniform magnification. Finally, with this solution, a distribution of non-uniform magnification is found, equalizing axial distances of focus of object spherical and parabolic waves.

Numerical and optical experiments accompany the paper. The numerical part proves NU-FDS algorithm capability for FoVs: 60° – 120° and resolutions: 2K – 16K. It consists of two parts, one examining a simple single-point object and the other investigating the CGHs reconstruction for two large 3D objects of complex shape. The first part carried out for 100° FoV confirms the high accuracy of reconstructing a large 2D target object. The second part shows that the method enables the reconstruction of large, highly detailed 3D objects. The numerical experiment presents the reconstruction of highly detailed CGH of 80° and 120° FoV. For smaller FoV, the optical experiment reconstructing the object of the same size and location is presented. The CGH of 120° FoV presents a numerical reconstruction of the object placed close to the observer with a height of more than 5m. For comparison, a reconstruction using the FrT method is also shown, which, yields very large shape deformations and incorrect depth of reconstruction.

**Funding.** Narodowe Centrum Nauki (UMO-2018/31/B/ST7/02980); Politechnika Warszawska.